\documentstyle[epsfig,12pt]{article}

\begin{document}
\setlength{\baselineskip}{5mm}

\noindent{\large\bf
Polynomial Heisenberg algebras and higher order supersymmetry
}\vspace{4mm}

\noindent{
\underline{David J. Fern\'andez C.}$^a$,
and V\'eronique Hussin$^b$
}\vspace{1mm}

\noindent{\small
$^a$Depto F\'{\i}sica, CINVESTAV, AP 14-740, 07000 M\'exico DF,
Mexico; $^b$D\'epartement de Math\'ema\-tiques, Universit\'e
de Montr\'eal, CP 6128, Succ Centre-Ville, Montr\'eal,
H3C 3J7 Canada
}\vspace{4mm}

\setlength{\baselineskip}{4.5mm}

\noindent {\small {\bf Abstract.} It is shown that the higher order
supersymmetric partners of the harmonic oscillator Hamiltonian provide the
simplest non-trivial realizations of the polynomial Heisenberg algebras. A
linearized version of the corresponding annihilation and creation operator
leads to a Fock representation which is the same as for the harmonic
oscillator Hamiltonian.}

\vspace{4mm}

\setlength{\baselineskip}{5mm}

\noindent {\bf 1. Introduction.} The non-linear algebras have attracted
attention in recent years due to the fact that they start to be applied in
physics. Those structures arise as deformations of a Lie algebra in which
some commutation relations are replaced by non-linear functions of the
generators [1-5]. Of particular interest are the simplest deformations of
the standard Heisenberg-Weyl algebra (the so-called polynomial Heisenberg
algebras) in which the commutator of the annihilation and creation
operators becomes a polynomial in the Hamiltonian but at the same time it
commutes with those ladder operators as in the standard case [6-11]. Let
us notice that these modified algebraic relations provide information
about the spectrum, which turns out to be a variant of the equally spaced
levels of the harmonic oscillator. Concrete realizations of those algebras
arise if the annihilation and creation operators become differential
operators of order greater than one [6,12]. The simplest non-trivial case
of such a realization is provided by considering the natural pair of
annihilation and creation operators for the Hamiltonians generated by
means of the higher order supersymmetric quantum mechanics (HSUSY QM) 
applied to the harmonic oscillator [11]. This will be the main subject
discussed in this paper. In order to do that, firstly we will present some
generalities of the polynomial Heisenberg algebras. Then, we will proceed
by discussing the standard SUSY and HSUSY QM as a mechanism to generate
solvable partner potentials from a given initial one [11,13-26]. From that
procedure it will be obvious the existence and the corresponding structure
of a pair of natural creation and annihilation operators for the HSUSY
partners of the oscillator potential, a very original construction
proposed by Mielnik for the first order SUSY [27-28]. It will be shown
that those operators realize in a simple way the polynomial Heisenberg
algebras. Later on it will be constructed a `linearized' version of those
algebras in which the modified ladder operators act onto the energy
eigenstates as the standard generators do in the Heisenberg-Weyl linear
case [11,29]. We will finish the paper with some general conclusions and
an outlook for future work. 

\bigskip

\noindent{\bf 2. Polynomial Heisenberg algebras.} Let us remember the
standard algebra of the harmonic oscillator:
\begin{equation}
[H,a] = -a, \quad [H,a^\dagger] = a^\dagger, 
\end{equation}
\begin{equation}
[a,a^\dagger] = 1.
\end{equation}
In the coordinates representation the operators $\{H, a, a^\dagger\}$ are
given by:
\begin{equation}
H = -\frac12\frac{d^2}{dx^2} + \frac12 x^2, 
\end{equation}
\begin{equation}
a = \frac{1}{\sqrt{2}}\left(\frac{d}{dx} + x\right),
\quad
a^\dagger = \frac{1}{\sqrt{2}}\left(-\frac{d}{dx} + x\right),
\end{equation}
the number operator reads:
\begin{equation}
N = a^\dagger a,
\end{equation}
and there is a {\it linear} dependence between $N$ and $H$: 
\begin{eqnarray}
& N = H - \frac12 \equiv N(H).
\end{eqnarray}

The polynomial Heisenberg algebras of $(n-1)$-th order are deformations of
the above algebra in which the two commutation relationships (1) are
maintained [6-11]:
\begin{equation}
[H,L] = - L, \quad [H,L^\dagger] = L^\dagger,
\end{equation}
but equation (2) is substituted by:
\begin{equation}
[L,L^\dagger] \equiv N(H+1) - N(H).
\end{equation}
The generalized number operator, defined by $N=L^\dagger L \equiv N(H)$,
becomes now a $n$-th order polynomial in the Hamiltonian $H$: 
\begin{equation}
N(H) = \prod_{i=1}^n \left(H - {\cal E}_i\right),
\end{equation}
and $L,\ L^\dagger$ are realized by $n$-th order differential operators. 
Let us notice that $[L,L^\dagger]$ is a $(n-1)$-th order polynomial in
$H$.

The algebraic properties of $\{H,L,L^\dagger\}$ provide already some
information about the spectrum of $H$. In order to see that, let us
analyze the space of solutions of the $n$-th order differential equation
(the kernel $K_{L}$ of $L$):
\begin{equation}
L\,\psi = 0 .
\end{equation}
Notice that $K_{L}$ is invariant under $H$. This and the fact that any
solution to (10) obeys: 
\begin{equation}
L^\dagger L\,\psi = \prod_{i=1}^n \left(H - {\cal E}_i\right)\psi = 0,
\end{equation}
suggest to select as the basis of $K_{L}$ those solutions which are
simultaneously eigenstates of $H$ with eigenvalues ${\cal E}_i$:
\begin{equation}
H\psi_{{\cal E}_i} = {\cal E}_i \psi_{{\cal E}_i} .
\end{equation}
However, some of the eigenfunctions $\psi_{{\cal E}_i}$ could be
non-normalizable. Let us suppose that $s$ of them, $\{\psi_{{\cal E}_i}, i
= 1,\dots,s\}$, are normalizable, which guarantee that they are
orthogonal. Taking those as {\it extremal} states one can construct, by
means of the iterated action of $L^\dagger$, $s$ energy ladders of
infinite length and spacing $\Delta E = 1$, each one of them starting from
${\cal E}_i$ (see Figure 1).

\begin{figure}[hp]\centering
\epsfig{file=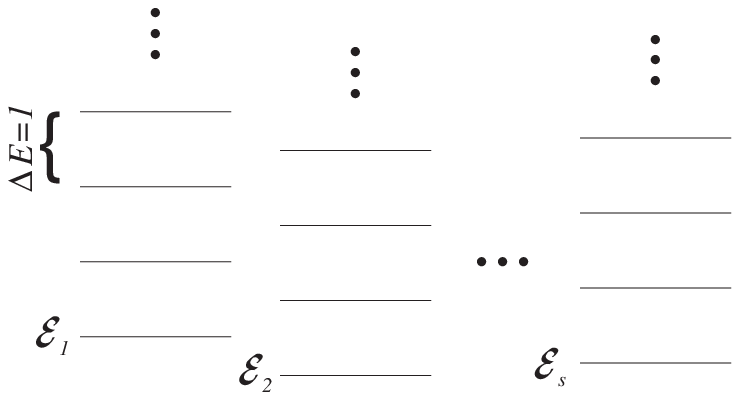, width=10cm}
\caption{\small Possible spectra for the Hamiltonians satisfying
equations (7-9). There are $s$ extremal states from which the $s$ infinite
ladders are constructed through the iterated action of $L^\dagger$.}
\end{figure}

It could happen, however, that for the ladder starting from ${\cal E}_j$
there exists a natural number $l$ such that: 
\begin{equation}
\left( L^\dagger\right)^{l-1}\psi_{{\cal E}_j} \neq 0, \quad \left(
L^\dagger\right)^{l}\psi_{{\cal E}_j} = 0.
\end{equation}
In such a case, by multiplying $(L^\dagger)^{l}\psi_{{\cal E}_j}$ to the
left by $L$, it is seen that one of the remaining roots of (11), $\{{\cal
E}_i, i=s+1,\dots,n\}$ is of kind ${\cal E}_k = {\cal E}_j + l$,
$k\in\{s+1,\dots,n\}, \ j\in \{1,\dots,s\}$. Thus, the spectrum of $H$
will consist of $s-1$ infinite ladders and a finite one of length $l$,
starting from ${\cal E}_j$ and ending at ${\cal E}_j+l-1$. An scheme of
this situation is represented in figure 2. This discussion means that the
$H$ of equations (7-9) is not, in general, the harmonic oscillator
Hamiltonian. 

\begin{figure}[htbp]\centering
\epsfig{file=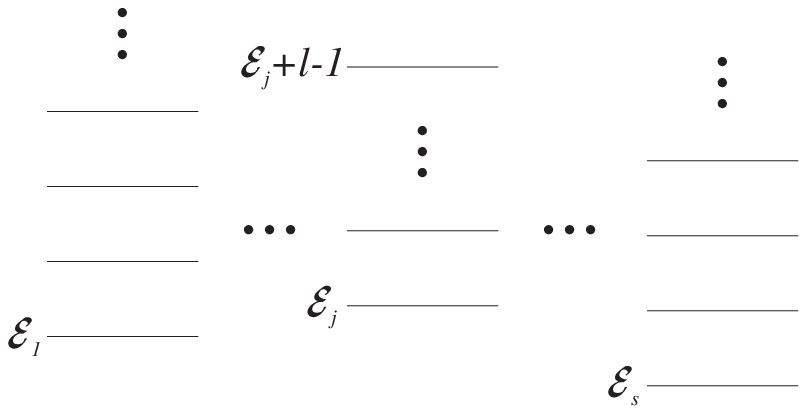, width=10cm}
\caption{\small The same as in figure 1 excepting that the $j$-th ladder
is finite due to the fact that $( L^\dagger)^{l}\psi_{{\cal E}_j} = 0$.  }
\end{figure}

\bigskip

\noindent{\bf 3. Standard supersymmetric quantum mechanics}.  The standard
SUSY algebra with two generators $Q_1, \ Q_2$, as introduced by Witten,
reads [13]: 
\begin{eqnarray}
&& [Q_i, H_{\rm ss}]=0, \nonumber \\ [-0.25cm] &&  \\ [-0.25cm] && \{
Q_i,Q_j\} = \delta_{ij} H_{\rm ss}. \nonumber
\end{eqnarray}
The simplest realization of such an algebra is of kind: 
\begin{eqnarray}
&& Q_1 =\frac1{\sqrt{2}}(Q^\dagger + Q),
\quad Q_2 = \frac1{i\sqrt{2}}(Q^\dagger - Q), \nonumber \\ [-0.25cm] &&
\\[-0.25cm]
&& Q =  \left(\matrix{ 0 & 0
\cr B & 0 }
\right), \quad Q^\dagger =
\left(\matrix{ 0 & B^\dagger \cr 0 & 0 } \right), \quad
H_{\rm ss} =
\left(\matrix{ B^\dagger B & 0 \cr 0 & B B^\dagger }
\right), \nonumber 
\end{eqnarray}
where usually $B^\dagger$ is a first order differential operator
\begin{equation}
B^\dagger \equiv A_1^\dagger = \frac{1}{\sqrt{2}}\left[-\frac{d}{dx} +
\alpha_1(x,\epsilon)\right], \quad B =
(B^\dagger)^\dagger,
\end{equation}
in such a way that $H_{\rm ss}$ is {\it linear} in $H^p = \left(\matrix{
\tilde
H & 0 \cr 0 & H } \right)$: 
\begin{equation}
H_{\rm ss} = (H^p-\epsilon).
\end{equation}
Notice that the Hamiltonians $H$, $\tilde H$ in the diagonal of $H^p$ are
intertwined by $B^\dagger$: 
$$
\tilde H B^\dagger = B^\dagger H,
$$
\vskip-0.6cm
\begin{equation}
H  =  - \frac{1}{2}\frac{d^2}{dx^2} + V(x),
\end{equation}
\vskip-0.3cm
$$
\tilde H  = - \frac{1}{2}\frac{d^2}{dx^2} + \tilde V(x). 
$$
This is equivalent to say that $V, \ \tilde V$ and $\alpha_1$ are related
through: 
\begin{eqnarray}
&& \alpha_1'(x,\epsilon)+
\alpha_1^2(x,\epsilon) = 2 [V(x)- \epsilon],
\nonumber \\[-0.25cm]  && \\[-0.25cm] && \tilde V(x) = V(x) -  
\alpha_1'(x,\epsilon).\nonumber 
\end{eqnarray}
The real number $\epsilon$ is called {\it factorization energy} (although
this does not mean that it should be a physical energy of $H$): 
\begin{eqnarray}
&& H = B B^\dagger + \epsilon, \nonumber \\[-0.25cm]  && \\[-0.25cm] &&
\tilde H
= B^\dagger B + \epsilon . \nonumber 
\end{eqnarray}
Let us denote by $\psi_n(x)$ the eigenfunctions of $H$, $H \psi_n(x) =
E_n \psi_n(x)$. From the intertwining relationship (18) it is simple to
get the eigenfunctions of $\tilde H$:
\begin{equation}
\tilde\psi_n(x) = \frac{B^\dagger
\psi_n(x)}{\sqrt{E_n - \epsilon}}, \quad \tilde H
\tilde\psi_n(x) = E_n \tilde\psi_n(x) .
\end{equation}
There is an eigenfunction $\tilde\psi_\epsilon(x)$ additional to those of
(21), which satisfies: 
\begin{equation}
B\tilde\psi_\epsilon(x) = 0 \quad \Rightarrow \quad
\tilde\psi_\epsilon(x) \propto
\exp\left[-\int_0^x\alpha_1(y,\epsilon)dy\right].
\end{equation}
The corresponding eigenvalue is $\epsilon$:
\begin{equation}
\tilde H \tilde\psi_\epsilon(x) = \epsilon \tilde\psi_\epsilon(x).
\end{equation}

We conclude that, departing from $H$, a new solvable Hamiltonian $\tilde
H$ with eigenfunctions $\{\tilde\psi_\epsilon(x), \tilde\psi_n(x)\}$ and
eigenvalues $\{ \epsilon,E_n\}$ can be generated. Let us remark the
importance of the value of $\epsilon$. It turns out that a necessary
condition in order to avoid supplementary singularities of $\tilde V(x)$
with respect to those of $V(x)$ is that $\epsilon$ is less than or equal
to the ground state energy $E_0$ of $H$. 

\bigskip

\noindent{\bf 4. Higher order supersymmetric quantum mechanics}. Let us
construct now a sequence of Hamiltonians $H_1, \dots, H_m$ departing from
$H_0$ as follows [11]:
\begin{eqnarray}
&& H_i A_i^\dagger  = A_i^\dagger H_{i-1},
\smallskip \nonumber \\ &&  H_i = -\frac12\frac{d^2}{dx^2} +
V_i(x), 
\\ &&  A_i^\dagger = \frac{1}{\sqrt{2}} \left[
-\frac{d}{dx} +
\alpha_i(x,\epsilon_i)\right]. \nonumber
\end{eqnarray}
Taking into account equations (18-19), we get once again a Riccati
equation and the corresponding expression of the potential at the $i$-th
step in terms of the $(i-1)$-th potential: 
\begin{eqnarray}
&& \alpha_i'(x,\epsilon_i) +
\alpha_i^2(x,\epsilon_i) = 2\left[V_{i-1}(x)  
-\epsilon_i\right], \nonumber \\[-0.25cm] && \\[-0.25cm] &&
V_i(x) = V_{i-1}(x) - \alpha_i'(x,\epsilon_i). \nonumber
\end{eqnarray}
The factorized expressions arising form the iterative procedure become:
\begin{eqnarray}
&& H_0 = A_1 A_1^\dagger + \epsilon_1 \nonumber \\ &&
H_i = A_i^\dagger A_i + \epsilon_i = A_{i+1}
A_{i+1}^\dagger + \epsilon_{i+1} \\ && 
H_m = A_m^\dagger A_m + \epsilon_m \nonumber
\end{eqnarray}
The corresponding eigenfunctions of $H_i$ are:
\begin{equation}
\cases{
\psi_{\epsilon_i}^{(i)}(x) \propto
\exp[- \int_0^x \alpha_i(y,
\epsilon_i) dy], \cr
\psi_{\epsilon_{i-1}}^{(i)}(x) =
\frac{A_i^\dagger \psi_{\epsilon_{i-1}}^{(i-1)}
(x)}{\sqrt{\epsilon_{i-1} - \epsilon_i}},
\medskip \cr  
\hskip2.0cm\vdots \medskip\cr
\psi_{\epsilon_1}^{(i)}(x) =
\frac{A_i^\dagger\dots A_2^\dagger
\psi_{\epsilon_1}^{(1)}(x)}{
\sqrt{(\epsilon_1-\epsilon_2)
\dots(\epsilon_1-\epsilon_i)}},
\cr
\psi_n^{(i)}(x) =  \frac{A_i^\dagger \dots
A_1^\dagger\psi_n^{(0)}(x)}{
\sqrt{(E_n-\epsilon_1)\dots(E_n -
\epsilon_i)}},}
\end{equation}
The eigenvalues become  $\{\epsilon_k, E_n, k=1,\dots,i,n=0,\dots\}$.

Let us make now in  (14-15) the following identifications:
\begin{eqnarray}
&& B=A_1\dots A_m, \nonumber \\
&& B^\dagger=A_m^\dagger\dots A_1^\dagger, \nonumber \\ [-0.25cm] &&
\\[-0.25cm] && 
H=H_0, \nonumber \\ &&
\tilde H = H_m. \nonumber
\end{eqnarray} 
With such a choice one obtains the higher order supersymmetric quantum
mechanics in which $H_{\rm ss}$ is a $m$-th order polynomial in $H^p$
[11,18-25]:
\begin{equation}
H_{\rm ss} = (H^p-\epsilon_1) \dots (H^p-\epsilon_m).
\end{equation}
By simplicity, we have ordered the factorization energies as $\epsilon_m <
\epsilon_{m-1}<\cdots < \epsilon_{1} \leq E_0$.  The final potential
$\tilde V(x)$ in terms of the initial one is given by: 
\begin{equation}
\tilde V(x) = V(x) -
\sum_{i=1}^m\alpha_i' (x,\epsilon_i).
\end{equation}
Thus, the point consists in looking for the solutions to the Riccati
equation (25) for the various values of $i$. It turns out that the $i$-th
superpotential $\alpha_{i}(x,\epsilon_{i})$ can be algebraically
determined from the $(i-1)$-th at $\epsilon_{i}$ and $\epsilon_{i-1}$
[22-25]: 
\begin{equation}
\alpha_{i}(x,\epsilon_{i}) = - \alpha_{i-1}(x,\epsilon_{i-1}) -
\frac{2(\epsilon_{i-1}- \epsilon_{i})}{\alpha_{i-1}(x,\epsilon_{i-1}) -
\alpha_{i-1}(x,\epsilon_{i})}.
\end{equation}
We arrive, finally, to the following conclusion: the $m$ solutions
$\alpha_1(x,\epsilon_i), i=1,\dots ,m$ to the initial Riccati equation
(19) determine completely, by means of just an algebraic composition
procedure involving equation (31), the final potential (30). 

\bigskip

\noindent{\bf 4.1. The harmonic oscillator.} By taking $V(x) = x^2/2$, it
is simple to find the general solution to (19) for an arbitrary fixed value
of $\epsilon$ [14,17]: 
\begin{eqnarray}
&& \nonumber \\ \alpha_1(x,\epsilon) = & -x + \frac{d}{dx}\bigg\{\ln
\bigg[
{}_1F_1\left(\frac{1-2\epsilon}{4},\frac12;x^2\right)  +
2\nu\frac{\Gamma(\frac{3 -
2\epsilon}{4})}{\Gamma(\frac{1-2\epsilon}{4})} \, x \,
{}_1F_1\left(\frac{3-2\epsilon}{4},\frac32;x^2\right)\bigg]\bigg\} .
\end{eqnarray}

In the first order supersymmetric quantum mechanics, we must have
$\vert\nu\vert <1$ in order to avoid singularities in $\tilde V(x)$. The
corresponding restriction for $\nu$ in the higher order case is, in
general, different from the one just pointed out [22]. The eigenfunctions
of the new potentials (30) are given by (27). The corresponding spectrum
takes the form $\{ \epsilon_i, E_n = n+1/2, i=1,\dots,m,n=0,1,\dots\}$,
i.e., there is one infinite ladder of equally spaced levels starting from
$1/2$ plus $m$ finite ones of length one and placed at arbitrary positions
$\epsilon_i, \ i=1,\dots,m$ below $1/2$. 

\newpage

\noindent{\bf 5. Polynomial Heisenberg algebras and HSUSY.} An obvious
constructions for a {\it natural} pair of annihilation and creation
operators associated to the HSUSY partners $\tilde H$ of the oscillator
can be seen from figure 3 [11,27-28]: 
\begin{equation}
D = B^\dagger a B, \qquad D^\dagger = B^\dagger a^\dagger B,
\end{equation} 
where $a$ and $a^\dagger$ are the annihilation and creation operators of
the harmonic oscillator. It is simple to show that $D, \ D^\dagger$ and
$\tilde H$ close a polynomial algebra of $(2m)$-th order of kind (7-9)
[11]: 
\begin{eqnarray}
&& [\tilde H,D] = -D, \nonumber \\ && [\tilde H,
D^\dagger] = D^\dagger, \\ && 
[D,D^\dagger]  =  N(\tilde H  + 1) - N(\tilde H). \nonumber
\end{eqnarray}
The generalized number operator $N(\tilde H)$ reads:
$$
 N(\tilde H) \!\equiv\! D^\dagger D \!=\!
\left( \tilde H \!-
\frac12\right)  
\prod_{i=1}^m \left( \tilde H \!-\! \epsilon_i \!-\! 1\right) \left(
\tilde H \!-\!
\epsilon_i \right).
$$

\begin{figure}[htbp]\centering
\epsfig{file=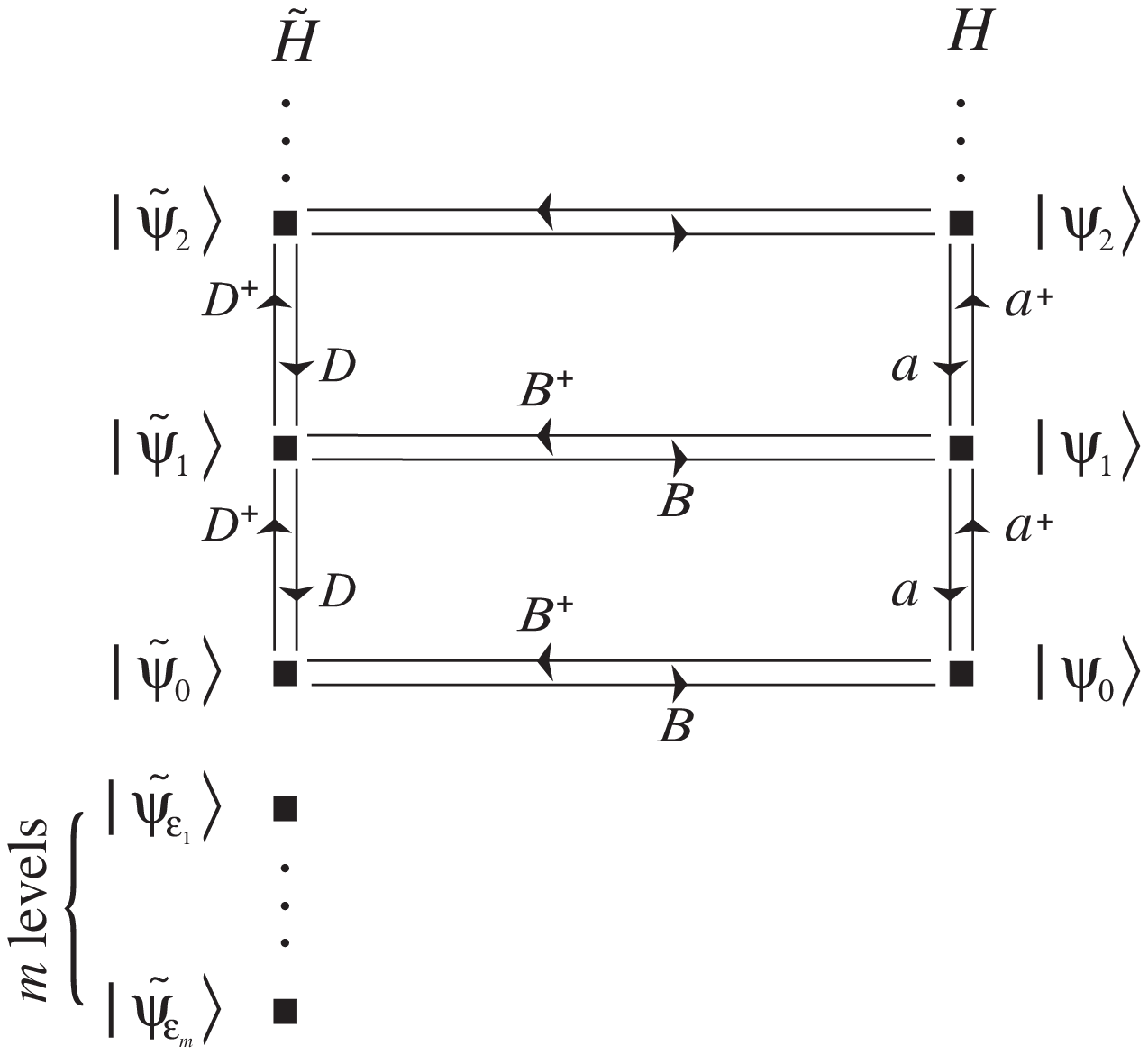, width=10cm}
\caption{\small The spectra of the HSUSY partners of the harmonic
oscillator. There is one infinite ladder starting from $E_0=1/2$ and $m$
finite ones starting and ending at $\epsilon_i, i=1,\dots,m$. }
\end{figure}

We notice that the eigenstates $\vert \tilde\psi_{\epsilon_i}\rangle$ of
$\tilde H$ are annihilated by both $D$ and $D^\dagger$. Moreover, any
eigenstate $\vert \tilde\psi_{n}\rangle, \ n=1,2, \dots$ of $\tilde H$ is
connected with its two neighbours $\vert\tilde\psi_{n-1}\rangle$ and
$\vert\tilde\psi_{n+1}\rangle$ by means of $D$ and $D^\dagger$ and the
extremal state $\vert \tilde\psi_{0}\rangle$ is just connected with
$\vert\tilde\psi_{1}\rangle$ by means of $D^\dagger$ because it is
annihilated by $D$. 

We have provided thus a non-trivial realization of the Polynomial
Heisenberg algebra (7-9). Next, we will look for a variant of the
annihilation and creation operators whose action onto the basis vectors
$\{\vert\tilde\psi_{\epsilon_i}\rangle, \vert\tilde\psi_{n}\rangle \}$
becomes similar as for the harmonic oscillator.

\newpage

\noindent{\bf 6. Linearized polynomial Heisenberg algebra.} The pair of
annihilation and creation operators $D_L$ and $D_L^\dagger$ acting onto
$\vert\tilde\psi_n\rangle, \ n=0,1,\dots$ as the Heisenberg-Weyl
generators take the form [11,29]: 
\begin{eqnarray}
&& D_L =  B^\dagger \left[
\prod_{i=1}^m (N - \epsilon_i +
\frac12)(N - \epsilon_i + \frac32)\right]^{-\frac12} a
B, \nonumber \\ [-0.25cm] && \\ [-0.25cm] &&
D_L^\dagger =  B^\dagger a^\dagger \left[
\prod_{i=1}^m
(N  -  \epsilon_i  + \frac12)(N 
- \epsilon_i  + \frac32)\right]^{-\frac12} B, \nonumber
\end{eqnarray}  
\begin{eqnarray}
&& D_L\vert\tilde\psi_n\rangle = \sqrt{n}
\vert\tilde \psi_{n-1}\rangle,
\nonumber \\ && D_L^\dagger\vert\tilde\psi_n\rangle =
\sqrt{n+1} \vert\tilde\psi_{n+1}\rangle, \\ && 
[D_L,D_L^\dagger]\vert\tilde\psi_n\rangle =
\vert\tilde\psi_n\rangle. \nonumber
\end{eqnarray}  
Let us notice that the expressions (35) for $D_L$ and $D_L^\dagger$ are
more involved than the corresponding ones for $D$ and $D^\dagger$ in (33). 
However, $D_L$ and $D_L^\dagger$ act onto $\{\vert \tilde\psi_n\rangle,
n=0,1,\dots\}$ simpler than $D$ and $D^\dagger$ do.  This property arises
again for the corresponding coherent states, a point which has been
discussed in detail elsewhere [11].

\bigskip 

\noindent{\bf 7. Conclusions and outlook.} The HSUSY QM is a powerful tool
in order to generate solvable potentials from a given initial one.  The
point reduces to find solutions to the initial Riccati equation (19) for a
set of factorization energies $\{\epsilon_i, \ i=1,\dots, m\}$. In the
oscillator case it is possible to iterate the process an arbitrary number
of times because it is known the general solution to (19) in the full
range $\epsilon < E_0$. The HSUSY partners of the oscillator have a pair
of natural annihilation and creation operators generating a $(2m)$-th
order polynomial Heisenberg algebra. This kind of algebras can be
linearized on the subspace spanned by the eigenvectors associated to $E_n
= n + \frac12, \ n=0,1,\dots$ It has been shown as well that $D_L$ and
$D_L^\dagger$ represent better than $D$ and $D^\dagger$ the role played by
$a$ and $a^\dagger$ for the oscillator potential. 

As a final remark, let us notice that the polynomial Heisenberg algebras
can be realized as well by the harmonic oscillator Hamiltonian and some
deformed versions of the standard annihilation and creation operators
[11]. This procedure can be considered artificial in the sense that a
complicated algebraic structure is constructed for a system which has from
the very beginning an inherently simpler algebra. We conclude that the
HSUSY partners of the oscillator provide the simplest systems for which
the polynomial Heisenberg algebras arise in a natural way. Let us notice,
however, that systems more general than the HSUSY partners of the
oscillator exist which have this kind of deformed algebraic structures
[6,12,26,30]. The study of such general systems is a subject of current
research which will continue in the near future. 

\bigskip

\noindent{Acknowledgments.} The authors acknowledge the support of CONACYT
(M\'exico), project 32086E. 

\newpage

\newcommand{\etal}{{\em et al.}}
\setlength{\parindent}{0mm}
\vspace{5mm}
{\bf References}
\begin{list}{}{\setlength{\topsep}{0mm}\setlength{\itemsep}{0mm}%
\setlength{\parsep}{0mm}}

\item[1.] M. Ro\v cek, Phys. Lett. B {\bf 255}, 554 (1991).

\item[2.] C. Daskaloyannis, J. Phys. A {\bf 24}, L789 (1991).

\item[3.] J. Beckers, Y. Brihaye and N. Debergh, J. Phys. A {\bf 32}, 2791
(1999).

\item[4.] C. Quesne and H. Vansteenkiste, math-ph/9908021; 
math-ph/0003025. 

\item[5.] V. Sunilkumar, B.A. Bambah, R. Jagannathan, P.K. Panigrahi and
V. Srinivasan, J. Opt. B {\bf 2}, 126 (2000).

\item[6.] S.Y. Dubov, V.M. Eleonsky and N.E. Kulagin, Sov. Phys.  JETP
{\bf 75}, 446 (1992). 

\item[7.] V.M. Eleonsky and V.G. Korolev, J. Phys. A {\bf 28}, 4973
(1995).

\item[8.] N. Aizawa and H.T. Sato, Prog. Theor. Phys. {\bf 98}, 707
(1997). 

\item[9.] F. Cannata, G. Junker and J. Trost, in {\it Particles, fields
and gravitation}, J. Rembielinski Ed., AIP Conf. Proc. {\bf 453}, Woodbury
(1998), p. 209. 

\item[10.] M. Arik, N.M. Atakishiyev and K.B. Wolf, J. Phys. A {\bf 32},
L371 (1999). 

\item[11.] D.J. Fern\'andez and V. Hussin, J. Phys. A {\bf 32}, 3603
(1999). 

\item[12.] D.J. Fern\'andez, M. Sc. Thesis, CINVESTAV (1984).

\item[13.] E. Witten, Nucl. Phys. B {\bf 188}, 513 (1981).

\item[14.] C.V. Sukumar, J. Phys. A {\bf 18}, 2917 (1985).

\item[15.] F. Cooper, A. Khare and U.  Sukhatme, Phys. Rep. {\bf 251}, 267
(1995).

\item[16.] D.J. Fern\'andez, J. Negro and M.A. del Olmo, Ann. Phys.  {\bf
252}, 386 (1996). 

\item[17.] G. Junker and P. Roy, Ann. Phys. {\bf 270}, 155 (1998). 

\item[18.] A.A. Andrianov, M.V. Ioffe, F. Cannata and J.P.  Dedonder, Int. 
J.  Mod. Phys. A {\bf 10}, 2683 (1995).

\item[19.] V.G. Bagrov and B.F. Samsonov, Phys. Part. Nucl. {\bf 28}, 374
(1997).

\item[20.] D.J. Fern\'andez, Int. J. Mod. Phys. A {\bf 12}, 171 (1997). 

\item[21.] D.J. Fern\'andez, M.L. Glasser and L.M. Nieto, Phys.  Lett. 
A {\bf 240}, 15 (1998). 

\item[22.] D.J. Fern\'andez, V. Hussin and B. Mielnik, Phys. Lett. A {\bf
244}, 309 (1998). 

\item[23.] J.O. Rosas-Ortiz, J. Phys. A {\bf 31}, L507 (1998); ibid 10163
(1998).

\item[24.] B.F. Samsonov, Phys. Lett. A {\bf 263}, 274 (1999). 

\item[25.] B. Mielnik, L.M. Nieto and O. Rosas-Ortiz, Phys. Lett. A {\bf
269}, 70 (2000). 

\item[26.] A. Andrianov, F. Cannata, M. Ioffe and D. Nishnianidze, Phys. 
Lett. A {\bf 266}, 341 (2000).

\item[27.] B. Mielnik, J. Math. Phys. {\bf 25}, 3387 (1984). 

\item[28.] D.J. Fern\'andez, V. Hussin and L.M. Nieto, J. Phys. A{\bf 27},
3547 (1994). 

\item[29.] D.J. Fern\'andez, L.M. Nieto and O. Rosas-Ortiz, J.  Phys. 
A{\bf 28}, 2693 (1995). 

\item[30.] A.P. Veselov and A.B. Shabat, Funct. Anal. Appl. {\bf 27}, 81
(1993). 

\end{list}

\end{document}